\documentclass[article,11pt]{IEEEtran}

\IEEEoverridecommandlockouts
\usepackage[utf8]{inputenc}
\usepackage[normalem]{ulem}
\usepackage{graphicx}
\usepackage{amstext}
\usepackage{amsmath}
\usepackage{amsfonts,amssymb}
\usepackage{amsmath,tabu}
\usepackage{amssymb}
\usepackage[dvips]{epsfig}
\usepackage{epsfig}
\usepackage[dvips]{graphicx}
\usepackage{multirow}
\usepackage{multirow} 
\usepackage[utf8]{inputenc}
\usepackage[english]{babel}
\usepackage{caption}
\usepackage{subcaption}
\usepackage{float}
\usepackage{booktabs}
\usepackage[nospace,noadjust]{cite}
\usepackage{algorithm,algorithmic}
\usepackage{fancyhdr}
\usepackage{marginnote}

\def\bth{{\boldsymbol{\theta}}}

\def\b1{{\boldsymbol{1}}}
\def\c1{{\textcircled{a}}}

\def\bb{{\boldsymbol{b}}}

\def\bn{{\boldsymbol{n}}}

\def\bq{{\boldsymbol{q}}}

\def\bs{{\boldsymbol{s}}}

\def\bx{{\boldsymbol{x}}}

\def\bz{{\boldsymbol{z}}}
\def\bA{{\mathbf{A}}}

\def\bC{{\boldsymbol{C}}}

\def\bF{{\mathbf{F}}}

\def\bI{{\mathbf{I}}}

\def\bS{{\boldsymbol{S}}}

\def\bZ{{\boldsymbol{Z}}}

\newtheorem{theorem}{Theorem}[section]

\newtheorem{lemma}[theorem]{Lemma}

\usepackage{xcolor}
\def\BibTeX{{\rm B\kern-.05em{\sc i\kern-.025em b}\kern-.08em
    T\kern-.1667em\lower.7ex\hbox{E}\kern-.125emX}}

\begin{document}
\bstctlcite{IEEEexample:BSTcontrol}
\title{Deep End-to-End Posterior ENergy (DEEPEN) for image recovery
}
\author{Jyothi Rikhab Chand, Member, IEEE, and Mathews Jacob, Fellow, IEEE
\thanks{The authors are with the Department of Electrical and Computer
Engineering, University of Virginia,
Charlottesville, VA 22904, USA (e-mail:
jyothi-rikhabchand@virginia.edu; mjacob@virginia.edu).
This work is supported by NIH R01AG067078, R01 EB019961, and R01 EB031169.}}

\maketitle

\begin{abstract}
Current end-to-end (E2E) and plug-and-play (PnP) image reconstruction algorithms approximate the maximum a posteriori (MAP) estimate but cannot offer sampling from the posterior distribution, like diffusion models. By contrast, it is challenging for diffusion models to be trained in an E2E fashion. This paper introduces a Deep End-to-End Posterior ENergy (DEEPEN) framework, which enables MAP estimation as well as  sampling. We learn the parameters of the posterior, which is the sum of the data consistency error and the negative log-prior distribution, using maximum likelihood optimization in an E2E fashion. The proposed approach does not require algorithm unrolling, and hence has a smaller computational and memory footprint than current E2E methods, while it does not require contraction constraints typically needed by current PnP methods. Our results demonstrate that DEEPEN offers improved performance than current E2E and PnP models in the MAP setting, while it also offers faster sampling compared to diffusion models. In addition, the learned energy-based model is observed to be more robust to changes in image acquisition settings. 
 
\end{abstract}

\begin{IEEEkeywords}
Energy model, MAP estimate, Memory-efficient, Parallel MRI, Uncertainty estimate
\end{IEEEkeywords}

\section{Introduction}
Computational algorithms that can recover images from sparse and noisy Fourier measurements have revolutionized magnetic resonance (MR) imaging. Traditional compressed sensing (CS) algorithms rely on hand-crafted image priors as regularizers \cite{tikhonov,donoho,candes}; wherein at each iteration they alternate between the update of data consistency (DC) and the proximal map of the regularizer to recover the unknown image. 

Data-driven deep learning methods have significantly improved reconstruction performance over classical methods. Plug-and-play (PnP) methods use an iterative algorithm that alternates between denoising and the DC update step, where the CNN-based pre-trained denoiser replaces the proximal map in the CS algorithm \cite{pnpbouman,consensus_equilibrium,rizwan_review}. A contraction constraint is often needed to ensure fixed-point convergence \cite{convergence}, which translates to reduced performance \cite{mol}. 
Recently, PnP methods that use explicit CNN-based energy functions to model the negative log-prior \cite{potential,gradientstep,MuSE} and are trained using denoising score matching (DSM) \cite{vincent2010} were introduced. These methods do not require a contraction constraint to guarantee stationary point convergence, translating into improved performance \cite{potential,gradientstep,MuSE}. Energy models also enable the generation of samples from the target distribution. 
The implicit multi-scale energy model (i-MuSE), which pre-learns a single energy function using DSM at multiple noise scales, was found to offer improved convergence and hence superior performance compared to a single-scale DSM \cite{MuSE}. Unlike PnP models that pre-learn CNN, end-to-end (E2E) methods \cite{lista,admmnet,variationalnet,modl,cinenet} unroll the iterative algorithm and optimize the parameters of the CNN denoiser so that the reconstructed image matches the reference image. These approaches offer better performance over PnP methods that are agnostic to the specific forward model. However, the E2E models are associated with increased memory demand, which can be reduced using the deep equilibrium (DEQ) formulation, but requires a contraction constraint similar to PnP methods that restricts its performance
\cite{mol, deq,deq_mri}. Since energy-based DEQ methods do not require this constraint \cite{elder}, they offer improved performance. 

The above approaches approximate the maximum a posteriori (MAP) estimate. In contrast, we learn the posterior distribution in an E2E fashion. Once trained, the learned posterior can be used to derive the MAP estimate and also sample from it. We derive the parameters of our Deep E2E Posterior ENergy (DEEPEN) network by maximizing the likelihood of data samples as in \cite{ebm}. In particular, we minimize the negative log-posterior of the training samples, which is the
sum of the negative log-likelihood and the negative log-prior
modeled by the CNN. The above training strategy simplifies to the minimization of the energy of the \emph{true} reference samples, while maximizing the energy of the \emph{fake} samples that are obtained using the Langevin sampling algorithm using the gradient of the learned posterior. The training strategy thus resembles a generative adversarial network (GAN) \cite{gan,mardanigan}. The key difference is that the generator involves a Langevin-based iterative algorithm specified by the energy model, and the classifier is also specified by the energy model. We show that when the Langevin noise level is small,  the above training strategy does not require algorithm unrolling or DEQ strategies associated with high memory demand and computational complexity. This E2E training strategy is computationally and memory-efficient because it does not require unrolling or fixed-point iterations. Moreover, unlike current E2E methods it does not use Mean Squared Error (MSE) loss to learn a MAP estimate. 

We note that the DSM approaches used in PnP models are trained to remove Gaussian noise perturbations, where the pixels of the perturbations are independent and identically distributed. 
However, the MR imaging inverse problem aims to recover images from corrupted undersampled Fourier measurements, where the corruption/perturbation often has highly correlated pixel values. Unfortunately, DSM-trained models
are often not efficient in estimating and removing correlated noise. By contrast, our experiments show that DEEPEN gradients are more effective in estimating Gaussian as well as correlated perturbations, thus offering improved performance in MAP image recovery. Our experiments also show that the learned E2E approach also generalizes to unseen acquisition settings, unlike traditional E2E approaches that are not robust to mismatch in forward models from the training settings. 

The DEEPEN model can also enable sampling from the posterior, where it can offer 10x faster sampling than diffusion models \cite{dps}, with comparable image quality, and reduced uncertainty, even when CNN complexity is 5x smaller. We note that deterministic ODE flows \cite{ddim,pnpflow} that learn straighter paths between probability distributions have been introduced to speed up prior sampling drastically. However, these models still require $>$100 steps in the posterior sampling setting. In particular, since they are trained only along the path between the distribution, noise often needs to be added after each gradient descent step involving data-consistency to ensure that the integration path remains in the regions trained originally  \cite{pnpflow}. 

A preliminary conference version of this approach was previously published with limited emperical results \cite{DEEPEN_ISBI}.
\section{End-to-end learned energy model }
Let $\bb \in \mathbb{C}^n$ denote the noisy undersampled measurements from which we wish to recover the unknown image $\bx \in \mathbb{C}^m$. The image and the measurements are related through a known linear forward operator $\bA \in \mathbb{C}^{n \times m}$:
\begin{equation}
    \bb = \bA \bx +\bn
\end{equation}
where $\bn \in \mathcal{N}(0,\eta^2 \bI) $ is additive complex white Gaussian noise. The recovery of the image from the measurements is often formulated as a MAP estimation problem, where the solution is the maximum of the log-posterior:
\begin{equation}\label{trueposterior}
    \log q(\bx|\bb) = \log p(\bb|\bx) + \log q(\bx)
\end{equation}
Here, $q(\bx)$ is the prior distribution, and the log-likelihood term $\log p(\bb|\bx)$ is specified by:
\begin{equation}\label{likelihood}
    \log p(\bb|\bx) = -\dfrac{\|\bA\bx-\bb\|^2}{2\eta^{2}} + P
\end{equation}
where $P$ is the normalization constant. The log-posterior can also be used to derive samples from the posterior distribution, which can inform about uncertainties in the estimation and offer likely solutions. 
\vspace{-3mm}
\subsection{Deep End-to-End Posterior ENergy (DEEPEN)}
We approximate the prior distribution $q(\bx)$ by a parametric energy model $p_\bth(\bx)$ as in \cite{ebm}:
\begin{equation}\label{prior}
    p_\bth(\bx) = \dfrac{1}{Z_\bth}\exp({-{\mathcal E}_\bth(\bx)}),
\end{equation}
where ${\mathcal E}_\bth(\bx): \mathbb{C}^m \rightarrow \mathbb{R}^+$ is modeled by a neural network and $Z_\bth$ is the normalization constant. 
Using \eqref{likelihood}, we obtain the parametric model for the log-posterior distribution as:
\begin{equation}\label{posterior}
    \log p_\bth(\bx|\bb) = \log p(\bb|\bx) + \log p_{\bth}(\bx),
\end{equation}
where $p_{\bth}(\bx)$ is the prior in \eqref{prior}.  Combining \eqref{likelihood} and \eqref{prior}, we obtain  the negative log-posterior $\mathcal L_{\boldsymbol\theta}(\bx)=-\log p_\bth(\bx|\bb)$ as:
\begin{eqnarray}
\label{eq:posterior_modeled}
\mathcal L_{\boldsymbol\theta}(\bx) &=&\underbrace{\dfrac{1}{2}\|\bA\bx-\bb\|^{2}+{\mathcal E}_\bth(\bx)}_{\mathcal C_{\bth}(\bx;\bA,\bb)} +\log \tilde{Z}_{\boldsymbol \theta},
\end{eqnarray}
where 
\begin{equation}
\label{normalizing}
    \tilde{Z}_{\boldsymbol \theta}=\int e^{-C_\bth(\bx;\bA,\bb)}d\bx
\end{equation}
 is a normalizing constant. For simplicity we absorbed the parameter $\eta^{2}$ into the definition of energy.\\
The common approach in energy-based models (EBMs) is to pre-learn $p_{\bth}$ from fully sampled images, independent of the forward operator $\bA$ \cite{ebm,msdm,MuSE}. Once the learning is complete, image recovery involves sampling from \eqref{posterior} using the
learned $p_{\bth}$ or maximizing \eqref{posterior} as in \cite{MuSE}. We note that learning in such PnP methods is agnostic to the forward model $\bA$. However, methods based on E2E training (for example, \cite{modl,variationalnet,mol,deq_mri}) often offer better performance than PnP methods. These E2E methods use pairs of measurements and images $(\bx,\bb)$ to learn $p_{\bth}$ so that the maximum of \eqref{posterior} matches the reference image, i.e., they try to learn the MAP estimate. Motivated by the improved performance of E2E methods, we propose to train the EBM in an E2E fashion for a specific forward operator $\bA$ in the next section.
\vspace{-3mm}
\subsection{Maximum likelihood training of DEEPEN}\label{ML loss}
We determine the optimal weights of ${\mathcal E}_\bth(\bx)$ by minimizing the negative log-likelihood of the training data set: 
\begin{eqnarray}\label{eq:KL}
\boldsymbol \theta^* &=& \arg \min_{\boldsymbol \theta}\:\mathbb{E}_{\bx \sim q(\bx)}\Big[\underbrace{-\log p_{\boldsymbol \theta}(\bx|\bb)}_{\mathcal L_{\boldsymbol\theta}(\bx)}\Big]
\end{eqnarray}
where the negative log-posterior $\mathcal{L}_{\boldsymbol\theta}(\bx)$ is specified by \eqref{eq:posterior_modeled}. 
Evaluating the gradient of the cost function in \eqref{eq:KL}, we obtain:
\begin{eqnarray}\nonumber
  \nabla_{\boldsymbol \theta}\mathcal{L}_{\boldsymbol \theta}(\boldsymbol x)&=& \mathbb{E}_{\bx \sim q(\bx)} [\nabla_{\boldsymbol \theta} {\mathcal C}_\bth(\bx;\bA,\bb)] + \nabla_{\boldsymbol \theta}  \log \tilde{\bZ}_{\boldsymbol \theta}\\\nonumber
  &=&\mathbb{E}_{\bx \sim q(\bx)} [\nabla_{\boldsymbol \theta} {\mathcal C}_\bth(\bx;\bA,\bb)]-\\&&\qquad  \mathbb{E}_{\bx \sim p_{\boldsymbol \theta}(\bx|\bb)}[\nabla_{\boldsymbol \theta} {\mathcal C}_\bth(\bx;\bA,\bb)].
\end{eqnarray}
In the second step, chain rule is used to simplify  $\nabla_{\boldsymbol \theta}  \log \tilde{\bZ}_{\boldsymbol \theta}$ using \eqref{normalizing} as in \cite{ebm}. Here,  $\bx \sim p_{\boldsymbol \theta}(\bx|\bb)$ are termed as \emph{fake samples}, drawn from the parametric posterior distribution $p_{\boldsymbol \theta}(\bx|\bb)$. For simplicity, we denote the fake samples as $\bx^-\sim p_{\theta}(\bx|\bb)$, while the reference samples are termed as \emph{true samples}, denoted by $\bx^+ \sim q(\bx)$. 

\begin{figure}[t!]
\centering
    \includegraphics[trim={4cm 9cm 5cm 6cm},clip,width=0.5\textwidth]{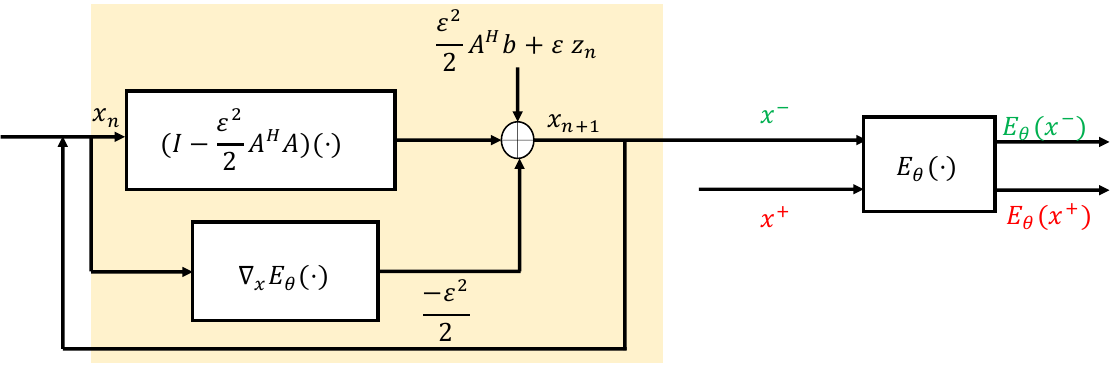}
    \caption{Demonstration of training procedure of DEEPEN. The training procedure determines the optimal weights of  the energy $E_{\boldsymbol \theta}(\cdot)$ by minimizing the energy difference between true and fake samples. The true samples $\bx^+$ are obtained from the training data, while the fake samples $\bx^-$ are generated using the Langevin sampling algorithm, highlighted by the yellow box. We note that the intermediate results are not stored to evaluate the loss's gradient; therefore, a single physical layer is used for forward propagation.  This keeps the training memory demand low. \vspace{-5mm}} \label{posterior_learning}
\end{figure}

Thus, the ML estimation of $\boldsymbol{\theta}$ is equivalent to minimization of the loss:
\begin{eqnarray}\nonumber\
 \mathcal{L}'(\boldsymbol \theta)&=& \mathbb{E}_{\bx \sim q(\bx)} {\mathcal C}_\bth(\bx;\bA,\bb) -\mathbb{E}_{\bx \sim p_{\boldsymbol \theta}(\bx|\bb)} {\mathcal C}_\bth(\bx;\bA,\bb)\\\nonumber\
  &\approx&  \left( \displaystyle \sum_{i=1}^{r}{{\mathcal C}_\bth(\bx^{+}_i;\bA,\bb)}-\displaystyle \sum_{j=1}^{r}{{\mathcal C}_\bth(\bx^{-}_j;\bA,\bb)}\right)\\\label{final}
\end{eqnarray}
where we consider $r$ samples. Intuitively, the training strategy (\ref{final}) seeks to decrease the energy of the true samples ${\mathcal C}_\bth(\bx^+;\bA,\bb)$ and increase the energy of the fake samples ${\mathcal C}_\bth(\bx^-;\bA,\bb)$. Thus, this approach may be seen as an adversarial training scheme similar to \cite{gan}, where the $\mathcal{C}_{\theta}$ serves as the classifier as shown in Fig. \ref{posterior_learning}. As in GAN models, the algorithm converges when the fake samples are identical in distribution to the training samples; i.e, ${\mathcal C}_\bth(\bx^+;\bA,\bb) \approx {\mathcal C}_\bth(\bx^-;\bA,\bb)$. Unlike GAN models that use a separate generator to create the fake samples, the production of the fake samples in the proposed approach also relies on $\mathcal C_{\theta},$ as described in the next subsection.
\vspace{-3mm}
\subsection{Generation of samples from posterior}\label{mcmc}
We generate the fake samples $\bx^- \sim p_{\boldsymbol \theta}(\bx|\bb)$ in \eqref{posterior} using Langevin Markov Chain Monte Carlo (MCMC)
method, which only requires the gradient of $\log p_{\boldsymbol \theta}(\bx|\bb)$ w.r.t. $\bx$:
\begin{eqnarray}\label{Ls}
\bx_{n+1}&=&\bx_n  - \dfrac{\epsilon^2}{2}  \nabla_{\bx }C_{\theta}(\bx_n;\bA,\bb)+\epsilon\bz_n \\
\nonumber
&=&\bx_n - \dfrac{\epsilon^2}{2} \left(\bA^H(\bA\bx_n-\bb)+\nabla_x {\mathcal E}_\bth(\bx_n)\right)+ \epsilon \bz_n 
\end{eqnarray}
where $\epsilon>0$ is the step-size, $\bz_n \sim \mathcal{N}(0,\bI)$, and $\bx_0$ is drawn randomly from a zero mean Gaussian distribution. Note that $\nabla_\bx {\mathcal E}_\bth(\bx)$ is the gradient of the energy model and does not depend on the normalization constant $\tilde{Z}_\bth$.

We now consider the gradient of  \eqref{final} w.r.t. $\bth$: 
\begin{eqnarray}\nonumber
\nabla_{\bth} \mathcal{L'}(\boldsymbol \theta)\approx 
\displaystyle \sum_{i=1}^{r}\nabla_{\bth}\,\mathcal C_{\bth}(\bx_i^+;\bA,\bb)-\displaystyle \sum_{j=1}^{r} {\nabla_{\bth}\,\mathcal C}_\bth(\bx^{-}_{j};\bA,\bb)\\\nonumber= \displaystyle \sum_{i=1}^{r}\nabla_{\bth}\,\mathcal E_{\bth}(\bx_i^+)-\displaystyle \sum_{j=1}^{r} {\nabla_{\bth}\,\mathcal E}_\bth(\bx^{-}_{j_{}})\\\nonumber -\sum_{j=1}^{r} \left(\nabla_{\bx} ~{\mathcal C}_{\bth}(\bx;\bA,\bb)|_{\bx^{-}_j}\right) \cdot \nabla_{\bth} ~(\bx^{-}_j)\\\label{final_unrolled}
\vspace{-7mm}
\end{eqnarray}
In the second step, we used the chain rule to expand $\nabla_{\bth}\,\mathcal E_{\bth}(\bx^-)$. In addition, we use the fact that the first term in \eqref{eq:posterior_modeled} is independent of $\bth$. We note that the last term in the second equation requires  unrolling of the Langevin iterations in \eqref{Ls}. However, we note that when $\epsilon\rightarrow 0$, the Langevin sampling in \eqref{Ls} simplifies to a gradient descent, which converges to the minimum of $ {\mathcal C}_{\bth}(\bx;\bA,\bb)$. Thus, we have $\left(\nabla_{\bx} ~{\mathcal C}_{\bth}(\bx;\bA,\bb)|_{\bx^{-}_j}\right)\approx\boldsymbol 0$ as $\epsilon \rightarrow 0$.  This implies that the last term in \eqref{final_unrolled} can be ignored, eliminating the need for unrolling. 

Therefore we obtain
 \begin{eqnarray}
     \nabla_\bth \mathcal{L}(\boldsymbol \theta)\approx \nabla_\bth {\mathcal E}_\bth(\bx^+) -\nabla_\bth {\mathcal E}_\bth(\bx_\bth^-)
 \end{eqnarray}
 which does not require backpropagation through the Langevin iterations, thus being efficient from a computational and memory perspective. The Langevin generation path gradient is usually not used in maximum likelihood training of EBMs \cite{ebm}. The training procedure is summarized in Fig. \ref{posterior_learning}.
\subsection{Image recovery}\label{map_algo}
Once training is complete, images can be derived from the posterior using Langevin sampling, specified by \eqref{Ls}. 

The proposed formulation also facilitates the estimation of MAP, where we minimize (\ref{eq:posterior_modeled}) w.r.t. $\bx$. In this work, we use the majorization minimization (MM) framework \cite{mm_tutorial} to derive the MAP estimate, which is guaranteed to converge monotonically \cite{mm_conv,mm_tutorial} to a minimum of \eqref{eq:posterior_modeled}. 
The MM framework consists of two steps \cite{mm_tutorial}. First, a surrogate function $g(\bx|\bx_n)$ is constructed such that $L_\bth(\bx) \leq g(\bx|\bx_n)$. Next, the surrogate function is minimized to get the next iterate. Similar to \cite{MuSE}, we used the following quadratic surrogate function to majorize $L_\theta(\bx)$ in (\ref{eq:posterior_modeled}):
\begin{eqnarray}\nonumber
    g(\bx|\bx_{n}) &=& \dfrac{\|\bA\bx - \bb\|^{2}}{2}+ {\mathcal E}_\bth(\bx_{n})+\dfrac{L }{2}\|\bx-\bx_{n}\|^{2}+\\\label{surrogate}
    &&\qquad\qquad\textrm{Re}\left(\boldsymbol H_{\theta}(\bx_{n})^{H}(\bx - \bx_{n})\right)
\end{eqnarray}
where $L$ is the Lipschitz constant of $\mathcal E_{\bth}$ and is approximately estimated using CLIP \cite{CLIP}. The above surrogate function has the following closed-form solution: 
\begin{equation}\label{mm_up}
        \bx_{n+1} = \left(\bA^{H}\bA+L\bI\right)^{-1}\left({\bA^{H}\bb} + L \bx_{n} - \boldsymbol H_{\theta}(\bx_{n})\right)
\end{equation}   
For a large-dimensional $\bx$, inverting the $\bA$ operator is computationally expensive. Therefore, we use the conjugate gradient algorithm to obtain the next iterate $\bx_{n+1}$.
The following result \cite{mm_conv} shows that the proposed algorithm converges monotonically to a stationary point of the cost function (\ref{eq:posterior_modeled}).
\begin{lemma}\label{l1}
Consider the cost function $\mathcal L_{\theta}(\bx)$ in (\ref{eq:posterior_modeled}), which is bounded below by zero\footnote{The CNN implementation $\mathcal{E}_\theta(\bx)$ has an absolute function in the output layer, which makes the lower bound zero.}. Then the sequence of iterates $\{\bx_n\}$ generated by MM algorithm in \eqref{mm_up} will converge to a stationary point of \eqref{eq:posterior_modeled}.
\begin{IEEEproof}
Note that $L_\bth(\bx) \geq 0; \forall \bx \in \mathbb{C}^m$ and hence it is lower bounded by a finite value. Moreover, the surrogate function satisfies $g(\bx_n|\bx_n) = f(\bx_n)$ and $g(\bx|\bx_n) \geq f(x)$. Then, using Theorem 1 from \cite{mm_conv}, the MM algorithm in \eqref{mm_up} will converge to a stationary point of \eqref{eq:posterior_modeled}.
\end{IEEEproof}
\end{lemma}

\section{MAP Experiments \& Results}
\subsection{Dataset} 
We compare the proposed DEEPEN approach with the SOTA methods described in the context of multichannel MR image reconstruction. The forward operator is defined as $\bA= \bS\bF\bC$, where $\bS$ is the sampling matrix, $\bF$ is the Fourier matrix, and $\bC$ is the Coil Sensitivity Map (CSM) that is estimated using \cite{espirit}.  MR images were obtained from the publicly available multichannel fastMRI brain dataset \cite{knee_dataset}. It is a 12-channel brain dataset and consists of complex images of size $320 \times 320$. The data set was divided into $45$ training, $5$ validation and $50$ test subjects, each consisting of approximately $450$, $50$, and $500$ images, respectively. We evaluated the models on T2-weighted images using $2D$ and $1D$ undersampling masks for different acceleration factors. We now describe the details of the proposed method as well as the SOTA methods.
\vspace{-2mm}
\subsection{Implementation details of the algorithms}
\subsubsection{DEEPEN}
We implement the energy model  $\mathcal E_\theta(\bx): \mathbb{C}^m \rightarrow \mathbb{R}^{+} $as a CNN consisting of five $3\times 3$ convolutional layers followed by a linear layer. Each convolutional layer consists of $64$ channels, and a Rectified Linear Unit (ReLU) was used between each layer, except the last linear layer. An absolute activation function was used in the linear layer to ensure that $\mathcal E_\theta(\bx)$ is lower bounded by a finite value. The score function $\nabla_\bx \mathcal E_\theta(\bx)$ was evaluated using the chain rule. 

To ensure stable ML training, similar to \cite{anatomy}, we found it useful to add zero-mean Gaussian noise with standard deviation $2\epsilon^2$ to the training data. We also found it beneficial to scale the posterior $p_\theta(\bx|\bb)$ by $\dfrac{\epsilon^2}{2}$, which gives rise to the equivalent Langevin MCMC update \cite{anatomy}:
\begin{equation}
\begin{array}{ll}\label{langevin_update}
\bx_{n+1}=\bx_n -  \left(\bA^H(\bA\bx_n-\bb)+\nabla_x {\mathcal E}_\bth(\bx_n)\right)+ \epsilon\bz_n 
\end{array}
\end{equation}
The MCMC Langevin algorithm was initialized with zero-mean Gaussian Noise and $100$ MCMC sampling iterations were performed to generate the fake samples. 
\subsubsection{PnP  MuSE}
We compare the proposed method with MuSE energy model that is trained in a PnP fashion. In particular, a single EBM is trained using DSM technique to predict Gaussian noise  corresponding to a range of noise standard deviations.  We used the following network architecture for MuSE:
\begin{eqnarray}
    E^{\rm{MuSE}}_\theta(\bx) = \dfrac{1}{2}\|\bx - \psi_\bth(\bx)\|^2
\end{eqnarray}
where $\psi_\bth(\bx)$ is a five layer convolutional network. Each layer was followed by ReLU activation, except in the final layer. The convolution layer consists of a $3 \times 3$ filter with 64 channels. MuSE was trained to predict Gaussian noise with standard deviations ranging from $0$ to $0.1$.
\subsubsection{PnP-ISTA}
PnP-ISTA is motivated by the iterative soft thresholding approach \cite{fista} used in CS algorithm. The proximal of the regularizer in the ISTA algorithm is replaced by a CNN denoiser, which is pre-trained as a Gaussian denoiser \cite{bekreview}. The optimization algorithm alternates between the following steps:
\begin{eqnarray}\label{ista}
    \bq_t = \bx_{t-1}- \alpha \dfrac{\bA^H(\bA \bx_{t-1}-b)}{\eta^2}\\
    \bx_t = D_\sigma(\bq_t)
\end{eqnarray}
where $D_\sigma(\cdot): \mathbb{C}^m \rightarrow \mathbb{C}^m$ is a five-layer CNN-based denoiser with each layer consisting of $64$ channels with a $3\times 3$ filter. Except for the final layer, after each layer the ReLU activation function was used. The denoiser was trained to remove Gaussian noise with standard deviation $\sigma =0.01$.  
\subsubsection{E2E MAP learning using CNN denoisers}
\begin{figure*}[htbp]
    \centering
    \begin{subfigure}[]{0.45\textwidth}
        \centering
        \includegraphics[width=1\textwidth]{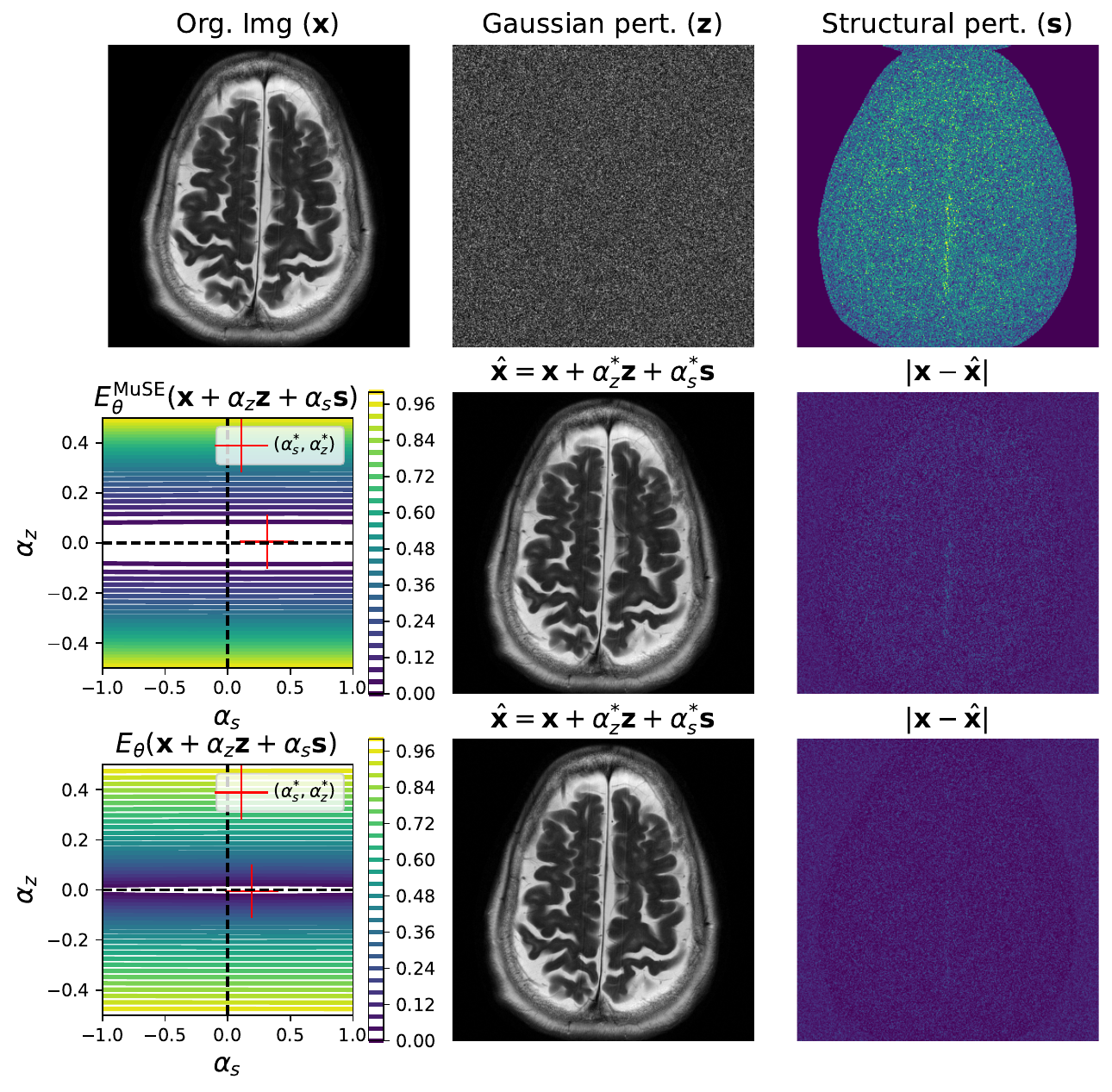}
        \caption{4-fold acceleration}
        \label{4-fold}
    \end{subfigure}
    \begin{subfigure}[]{0.45\textwidth}
        \centering
        \includegraphics[width=1\linewidth]{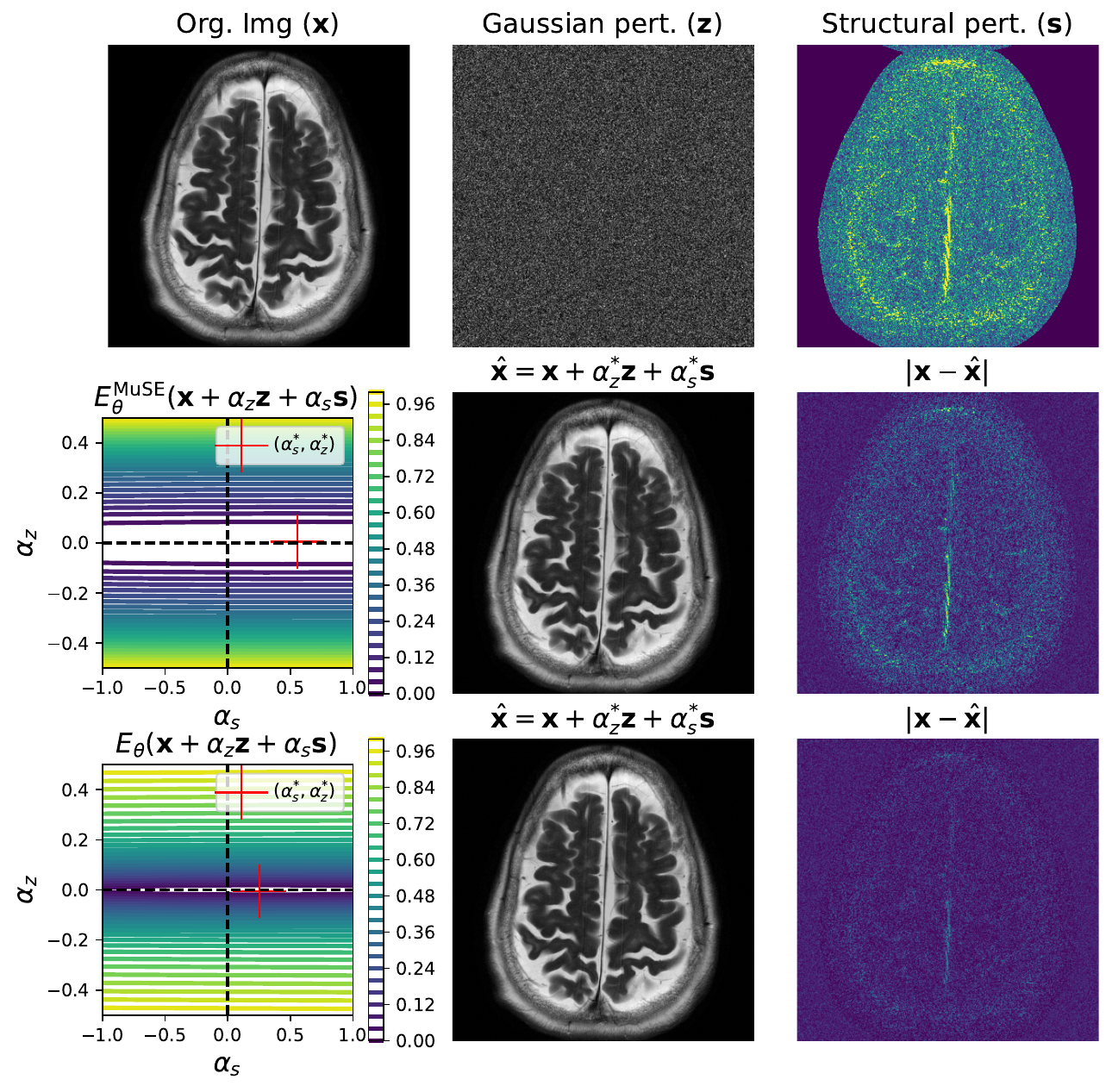}
        \caption{6-fold acceleration}
        \label{6-fold}
    \end{subfigure}
    \caption{Comparison of pre-trained (MuSE) with E2E-trained (DEEPEN) energy models for (a) four-fold and (b) six-fold acceleration. The top row in each figure shows the original image, Gaussian noise, and the structural perturbation specified by $\bx_{0}-\bx$, where $\bx_0$ denotes the sense solution. The second and third row in each figure shows the plot of MuSE and DEEPEN energy as a function of $\alpha_z$ and $\alpha_s$, their corresponding reconstructed and the error images, respectively. The images are reconstructed by taking the combination of the form $\hat{\bx} = \bx + \alpha_s^{*}\bs + \alpha_z^{*}\bz$, where $(\alpha_s^{*},\alpha_z^{*})$ are the minimizer (indicated by cross mark in the contour plot) of the energy function. We note that the minimum of the DEEPEN energy is closer to $\alpha_s^{*}\approx 0;\alpha_z^{*}=0$, with the differences $\bx^*-\bx$ smaller than that of MuSE. This shows that the DEEPEN energy is effective in suppressing both correlated structural perturbations as well as Gaussian noise. }
    \label{energy_illustration}
\end{figure*}
We also compare EBMs with an E2E approach MoL \cite{mol}, which approximate the MAP estimate and are trained to minimize the MSE between the recovered images and the reference images. 
\begin{eqnarray}\label{MSE}
    L_{\rm{MSE}}(\bth) = \displaystyle\sum_{i=1}^{N_{\rm samples}}\|\bx^*_{\bth}(i) - \bx_{\rm ref}(i)\|^{2}
\end{eqnarray}
Here, $N_{\rm samples}$ denotes the number of training samples,  $\bx_{\rm ref}(i)$ is the reference image, and  $\bx_{\bth}^*(i)$ is the solution to a regularized optimization problem that uses the CNN as a regularizer \cite{mol}. The gradient of \eqref{MSE} is specified by 
\begin{eqnarray}
    \nabla_{\bth}~L_{\rm{MSE}}(\bth) = \sum_{i=1}^{N_{\rm samples}}\underbrace{(\bx^*_{\bth}(i) - \bx_{\rm ref}(i))}_{l_i}\cdot \nabla_{\theta} \bx^*_{\bth}
\end{eqnarray}
Unlike the last term in \eqref{final_unrolled}, $l_i$ is not zero in the above case. The DEQ  approach \cite{deq,deq_mri,mol} is used to derive the above gradient, which assumes the iterative algorithm to derive $\bx^*_{\bth}$ to converge to a fixed point. A monotone constraint is placed on the CNN to ensure convergence to the fixed point. We used a five-layer CNN to implement MoL and the monotone constraint was imposed using a log-barrier approach as described in \cite{mol}.

\subsubsection{E2E MAP using CNN energy models}\label{MSE loss}
The DEEPEN framework learns the posterior distribution using \eqref{eq:KL} in an E2E fashion, which can be used for posterior sampling or to derive MAP estimates. In contrast, ELDER in \cite{elder} learns the regularizer parameters in an E2E fashion by minimizing the loss of MSE in \eqref{MSE}, similar to the MoL approach described in the above section. In this work, we consider  the following iterative algorithm for $K$ iterations: 
\begin{eqnarray}\label{gd_up}
    \bx_{{k+1} }= \bx_{{k} } - \alpha\left(\bA^H(\bA\bx_k-\bb)+\nabla_{\bx_{k}} \mathcal E_{\bth}(\bx_{{k} })\right)
\end{eqnarray}
to obtain the MAP estimate. Here, $\alpha$ is the step-size and is a learnable parameter. 
\subsection{Visualization of MuSE and DEEPEN energies}
In this section, we compare pre-trained MuSE with DEEPEN, whose energy model is trained in E2E fashion. For both energy models, we fed images of the form $\tilde{\bx} = \bx + \alpha_z\bz + \alpha_s \bs$, where $\bx$ is the test image, $\bz \sim \mathcal{N}(0, \bI)$ is the Gaussian noise perturbation, $\bs = (\bx_0-\bx) $ is the structural artifact perturbation and $\bx_0$ is the SENSE solution given as $\bx_0 = (\bA^H\bA+ \tilde{\lambda}\bI)^{-1}\bA^H\bb$. An example of the test image, Gaussian and structural perturbation is shown in the top rows of Fig.\ref{energy_illustration}.a and Fig.\ref{energy_illustration}.b for four-fold and six-fold acceleration, respectively.  The second and third rows of each figure show the MuSE and DEEPEN energy plot as a function of $\alpha_s$ and $\alpha_z$ for four-fold and six-fold acceleration, respectively. The red cross on the plot indicates $(\alpha_s^{*},\alpha_z^{*})$ for which the energy function achieves a minimum value. Note that, a well-trained energy function will have a lower energy value for good images. 

We observe from the figure that unlike MuSE, the energy model trained via DEEPEN has $(\alpha_s^{*},\alpha_z^{*})$ closer to zero, for both acceleration settings. This implies that the DEEPEN energy model is a better discriminator between the reference and the image with correlated perturbations. We also show the image $\hat{\bx} = \bx + \alpha_s^{*}\bs + \alpha_z^{*}\bz$ at minimum in the second and third rows of Fig.\ref{energy_illustration} and the corresponding error images. We observe that DEEPEN offers an image that is closer to the reference image.  The difference between the minimizer and reference can be appreciated from the error images, especially for six-fold acceleration, where the error image of MuSE has a significant structural perturbation compared to DEEPEN.  This shows that DEEPEN is a better regularizer and, consequently, when employed in inverse problems, will be efficient in removing structural perturbations.

\begin{table*}[h!]
\centering
\caption{Reconstruction performance of different models  for four different acquisition settings.}
\begin{tabular}{|l|ll|ll|ll|ll|}
\hline
\multirow{2}{*}{Algorithm} & \multicolumn{2}{c|}{2D 4x Mask} & \multicolumn{2}{c|}{2D 6x Mask} & \multicolumn{2}{c|}{1D 2x Mask} & \multicolumn{2}{c|}{1D 4x Mask} \\ \cline{2-9} 
                           & Avg. PSNR         & SSIM        & Avg. PSNR         & SSIM        & Avg. PSNR         & SSIM        & Avg. PSNR         & SSIM        \\ \hline
DEEPEN                          &   39.15+/-1.43              &       0.98         &   37.18+/-1.32                & 0.975             &    38.91+/-2.21               &  0.98           &  31.24+/-1.64               &    0.93         \\
MuSE               & 38.27+/-1.30                  &    0.97         &      36.64+/-1.20             &   0.96          &    38.37+/-1.88               & 0.97            &   31.66+/-1.75            &   0.93          \\ 
ISTA  &    39.88+/-1.38               &  0.97           &   36.76+/-5.15                &  0.96           &               39.91+/-2.93    & 0.98            &   29.04+/-10.80                &  0.93           \\ 
MoL &              38.12+/-1.27     &   0.97          &   36.91+/-1.19                &     0.97        &  37.72 +/-1.67                 &    0.97         &            31.35+/-1.38       &   0.926          \\ 
ELDER &  38.66+/-1.23                 &      0.98       &    37.39+/-1.17               &    0.97         & 38.02+/-1.52                  &  0.97           &  31.19+/-1.21                 & 0.93            \\ 

\hline
\end{tabular}
\label{recon_cmp}
\end{table*}

\begin{figure}[t]
    \centering
    \begin{subfigure}[b]{0.49\textwidth}  
        \includegraphics[width=1\linewidth]{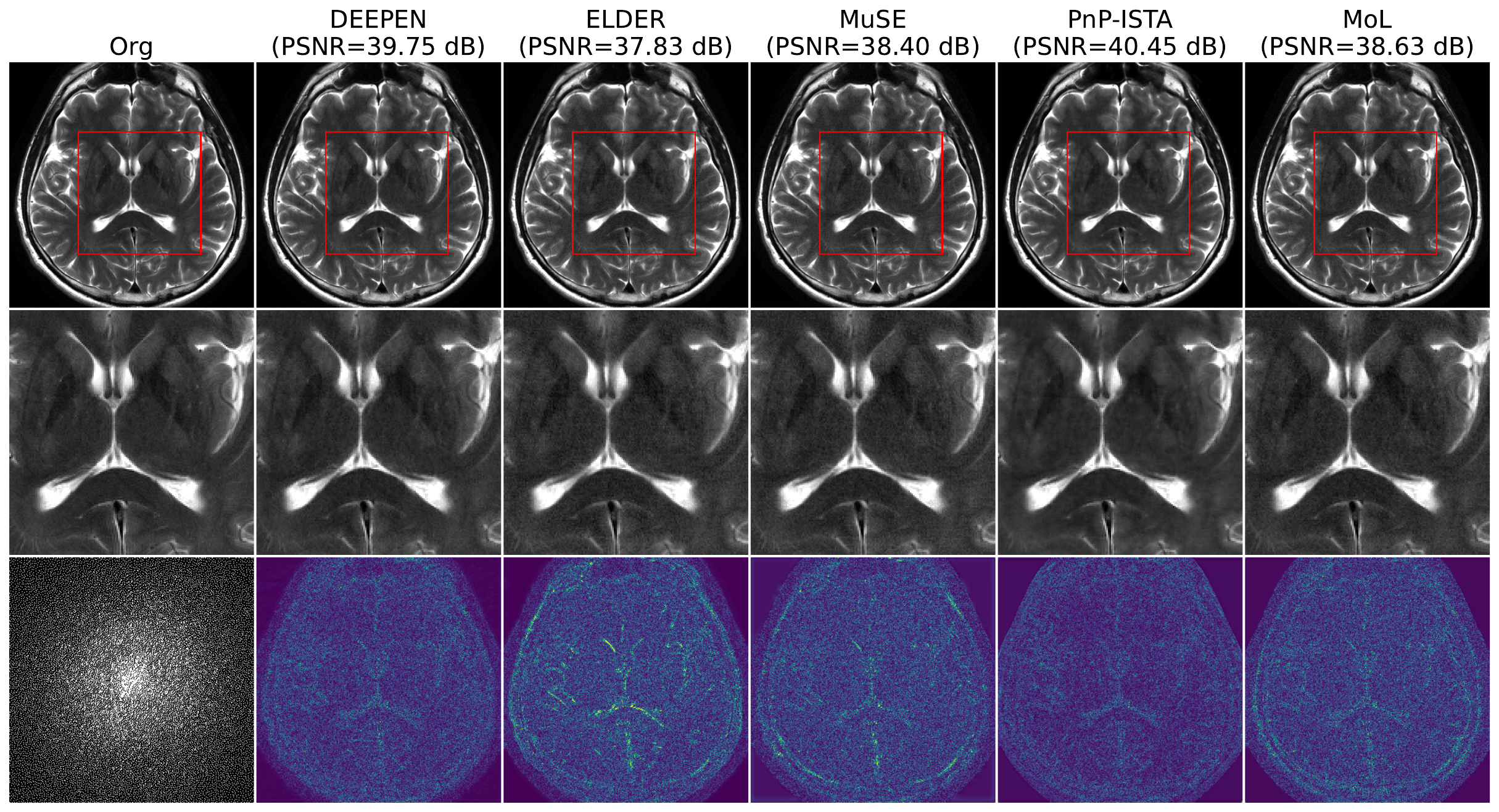}
    \caption{Four-fold acceleration using 2D undersampling mask}
    \label{four_fold}
    \end{subfigure}
       \begin{subfigure}[b]{0.49\textwidth} 
        \includegraphics[width=1\linewidth]{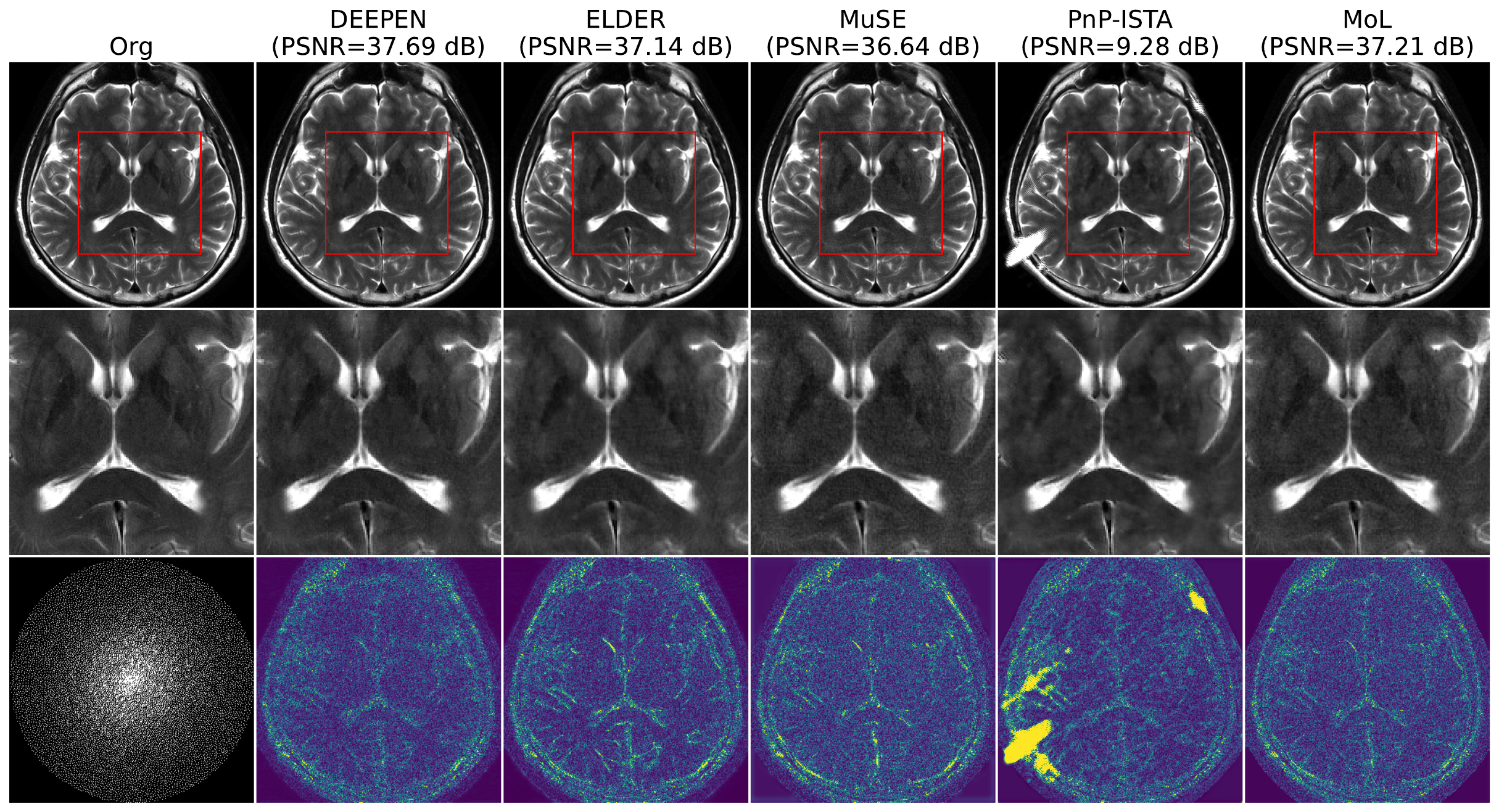}
    \caption{Six-fold acceleration using 2D undersampling mask}
    \label{six_fold}
    \end{subfigure} 
     \begin{subfigure}[b]{0.49\textwidth} 
        \includegraphics[width=1\linewidth]{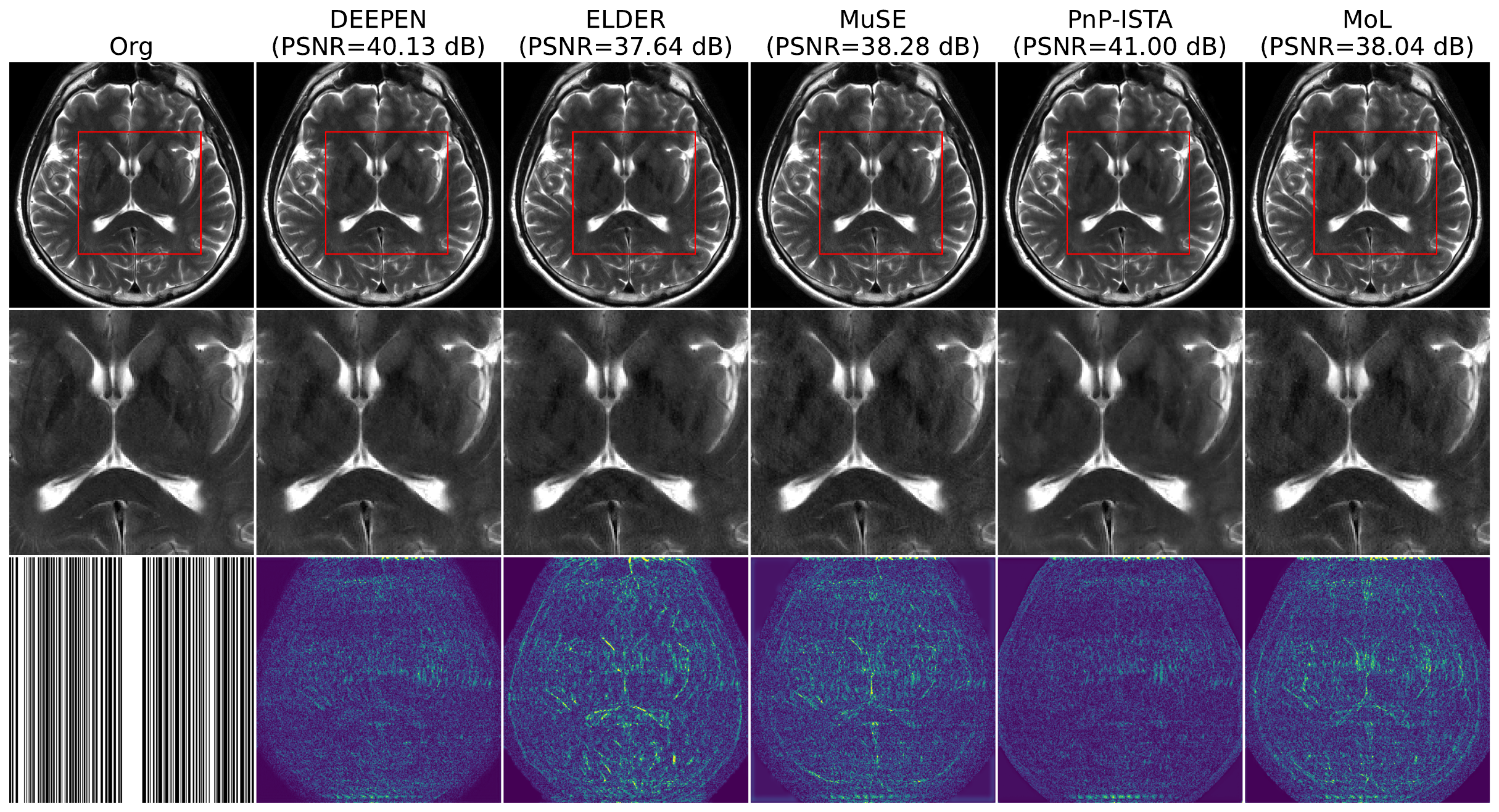}
    \caption{Two-fold acceleration using 1D undersampling mask}
    \label{two_fold_1D}
    \end{subfigure} 
    \caption{Comparison of DEEPEN, ELDER, MuSE, PnP-ISTA, and MoL for three different acquisition settings on the fastMRI brain data set. The first, second, and the third row in each figure shows the reconstructed image, enlarged image, and the error image, respectively. The error image is scaled by a factor of 10 to highlight the differences. The first image in the third row shows the undersampling mask.\vspace{-8mm} } \label{mse_cmp}
\end{figure}
\subsection{Reconstruction}
In this section, we compare the reconstruction performance of DEEPEN with MuSE, ELDER, PnP-ISTA, and MoL. The energy-based models DEEPEN and MuSE were run until the cost function satisfied $|L_\theta (\bx_{n+1}) - L_\theta(\bx_{n})|/ |L_\theta(\bx_{n}) | \leq 1e^{-6}$ or until $500$ iterations were reached. The optimal hyperparameters needed to run MuSE were obtained from \cite{MuSE}. We use the update equation in \eqref{gd_up} as the inference algorithm for ELDER trained with MSE loss with $K=30$. PnP-ISTA was run until $|\bx_{n+1} - \bx_{n}|/ |\bx_{n} | \leq 1e^{-6}$ or until $500$ iterations were reached. \\
 All reconstruction algorithms were initialized with SENSE. Table. \ref{recon_cmp} compares the reconstruction performance for four different settings: 4x and 6x acceleration with 2D undersampling mask; 2x and 4x acceleration with 1D undersampling mask. The reconstructions are compared using two metrics: Peak Signal-to-Noise Ratio (PSNR) and Structural Similarity Index Measure (SSIM). From the table, we observe that DEEPEN offers better results compared to MuSE. The improved performance of DEEPEN can be attributed to the training strategy, which ensures that the true samples correspond to the minimum of energy, with it increases in other directions. In particular, the energy is higher for both Gaussian and structural perturbations (as demonstrated in Fig. \ref{energy_illustration}), compared to MuSE.  However, we note that both energy models learn a negative log-prior distribution that ensures convergence (without a contraction constraint on the score function), as discussed in Lemma \ref{l1}. We note that the performance of DEEPEN is comparable to that of PnP-ISTA at lower undersampling rates (e.g. 2D mask 4-fold setting). PnP-ISTA was implemented without contraction constraints, which offers higher performance than the implementation with constraints \cite{MuSE}. We also observe that the performance of PnP-ISTA drops significantly at higher accelerations, which is consistent with the observation in \cite{MuSE}. In particular, PnP-ISTA exhibits localized artifacts (bright regions in Fig. \ref{mse_cmp}.(b)) due to convergence problems. We observe that DEEPEN offers improved performance than ELDER for the 2D 4-fold and 1D 2-fold acquisition settings, while the performance is comparable for other settings. 
 \vspace{-5mm}
\subsection{Generalization performance of E2E energy models}
\begin{figure}[h]
    \centering
    \begin{subfigure}[]{0.4\textwidth}
        \centering
        \includegraphics[width=0.8\linewidth]{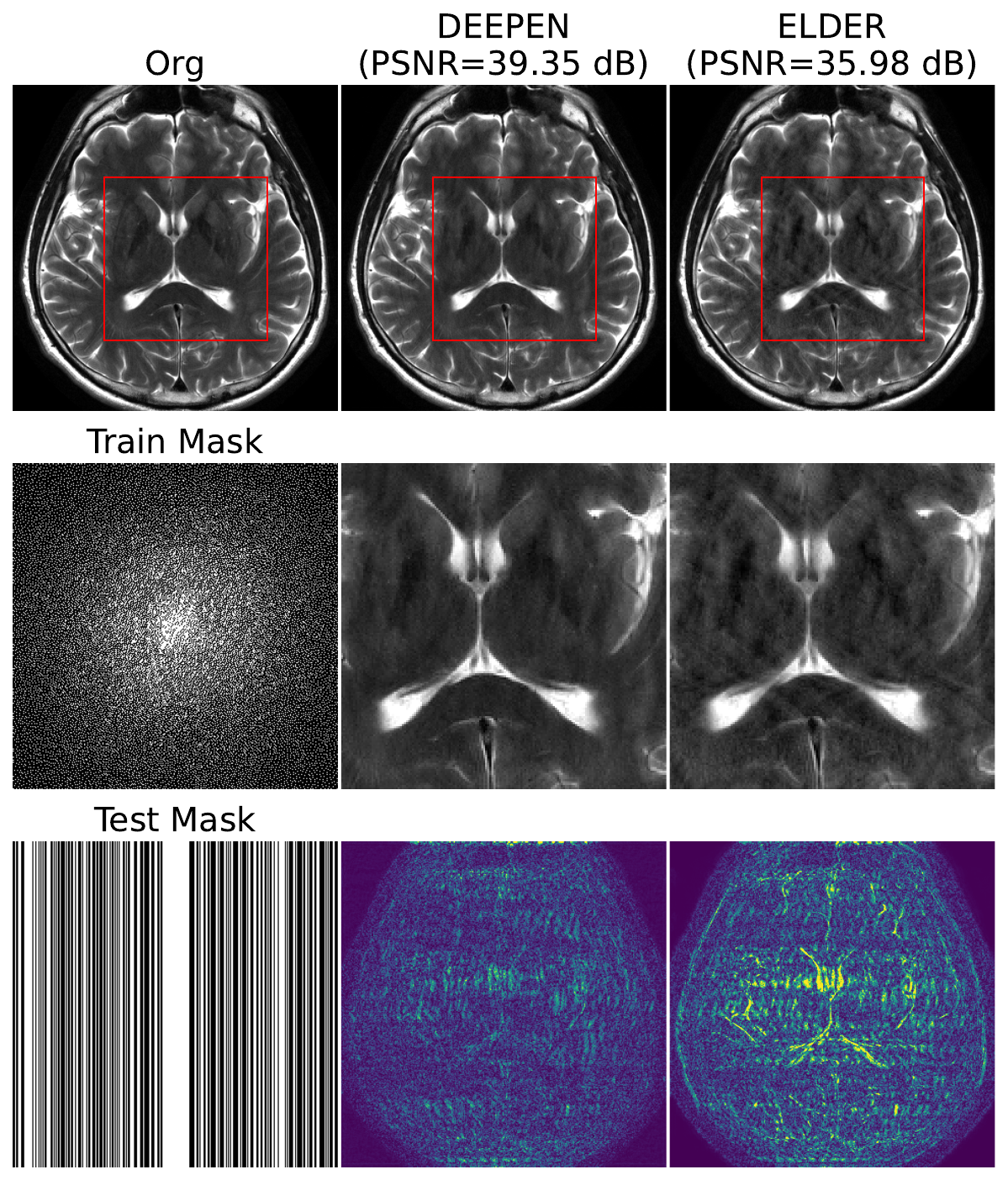}
        \caption{}
    \end{subfigure}
    \begin{subfigure}[]{0.4\textwidth}
        \centering
        \includegraphics[width=0.8\linewidth]{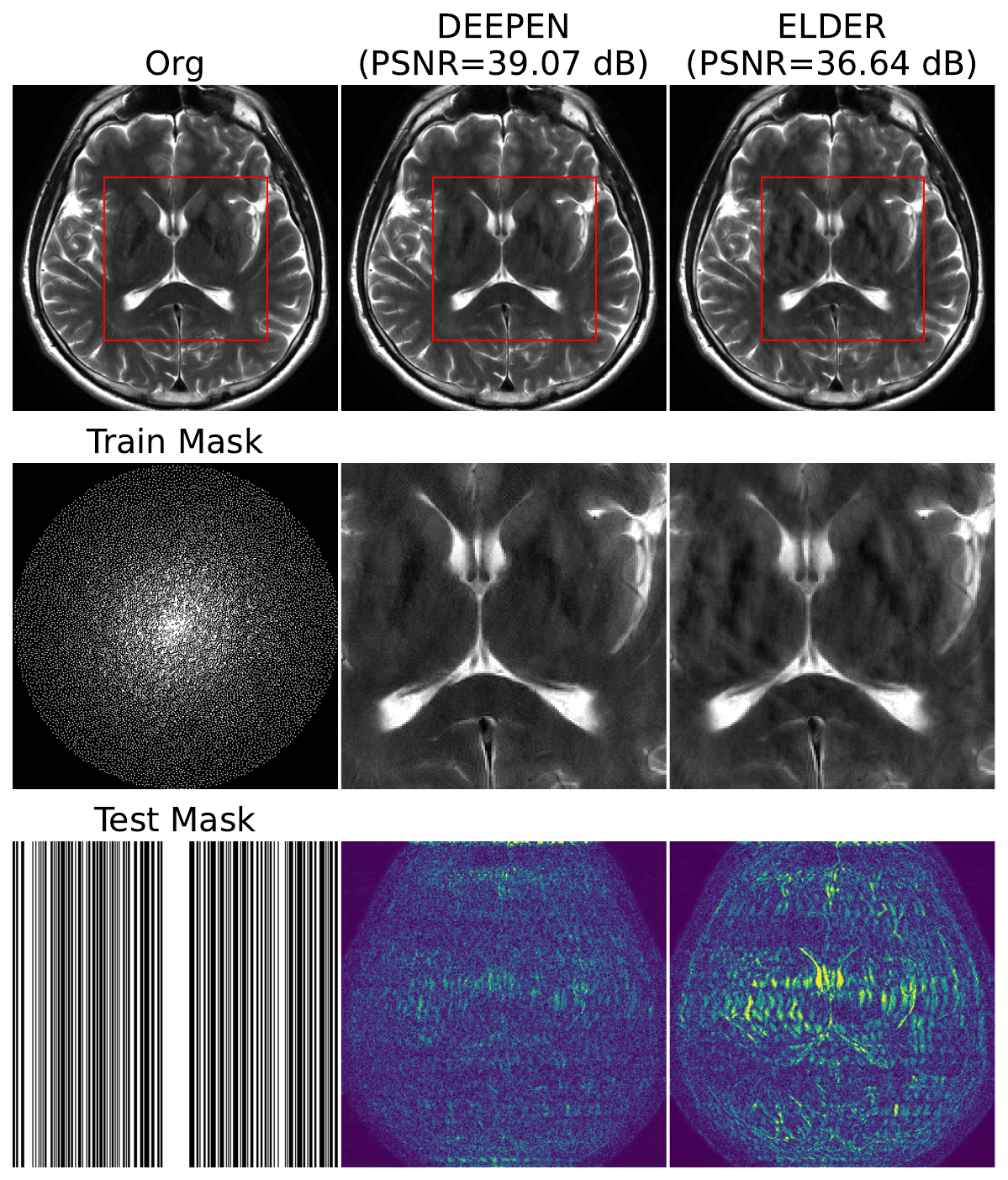}
        \caption{}
    \end{subfigure}
  \caption{Generalization comparison of E2E-trained DEEPEN and ELDER models for two different settings: (a) models are trained on 4-fold 2D undersampling mask and tested on a 2-fold 1D undersampling mask. When compared to Fig. \ref{mse_cmp}.a, which demonstrates the reconstruction performance of models trained and tested on 4-fold 2D mask, the performance of DEEPEN and ELDER drops by $0.4$ dB and $1.85$ dB, respectively  (b) models are trained on 6-fold 2D mask and tested on 2-fold 1D mask. When compared to Fig. \ref{mse_cmp}.b, which demonstrates the performance of models trained and tested on 6-fold 2D mask, the performance of DEEPEN improved by $1.38$ dB while the performance of ELDER dropped by $0.5$ dB.\vspace{-5.5mm} }
   \label{generalization_fig}
\end{figure}

DEEPEN and ELDER both use energy models, which are trained in E2E fashion. The main difference is that DEEPEN learns the posterior using the maximum likelihood approach, while ELDER relies on MSE loss. We compare the generalization performance by considering a different acquisition scheme (sampling pattern) during the test setting from what was assumed during training. Table \ref{generalization} compares the performance of DEEPEN and ELDER for different acquisitions. In the table, we report the Avg. PSNR, where we have boldfaced the performance of DEEPEN. From the table, we observe that when there is a change in the acquisition setting, the performance of ELDER drops drastically compared to DEEPEN. For example, when the energy models are trained on a 4-fold 2D undersampling mask and tested in the same setting, the performance of DEEPEN and ELDER is about $39.15$ dB and $38.66$ dB, respectively. However, when tested on a 2-fold 1D mask, the performance of ELDER drops by about $2.38$ dB while the performance of DEEPEN drops only by $0.92$ dB. Fig. \ref{generalization_fig} illustrates the generalization performance of DEEPEN for two different mismatch scenarios.

We attribute the improved generalization performance of DEEPEN to the adversarial training strategy. We note that the fake samples are generated by Langevin dynamics, where Gaussian noise is added at every iteration to generate the fake samples. This training strategy ensures that the energy function is well-trained along different random paths from the initialization. This approach also encourages the energy function to have well-defined minima in the training samples, with the energy values increasing in all directions, as seen in Fig. \ref{energy_illustration}. By contrast, training with the MSE loss learns a single path from the initialization and the reference image. We note that the iterations during inference depend on the energy function as well as the forward model; a change in the forward model can result in a different path to the samples, which may make E2E methods trained using MSE loss less robust to changes in acquisition settings.  
\section{Posterior Sampling Experiments}
Most of the current E2E deep learning methods focus on learning the MAP estimate \cite{modl,variationalnet,mol,elder}. In contrast, DEEPEN learns the posterior distribution in an E2E fashion, which enables us to sample the distribution using \eqref{langevin_update}. We note that the long sampling chain in diffusion models and the time dependence of the score model make it challenging to train diffusion models in an E2E fashion. We compare the sampling performance of DEEPEN with Diffusion Posterior Sampling (DPS) \cite{dps}, where the diffusion model is learned in a PnP fashion in the T2-weighted fastMRI brain dataset.  We also note that it is challenging for diffusion models to realize MAP estimate. In particular, score models requires the computation of the following  complex integral to estimate the log-prior \cite{yansong}:
\begin{equation}\label{LineInt_diffusion}
    \log p_\theta(\bx) = \int_{0}^{T}\nabla .{s}_\theta (\bx(t),t) dt+ \log p_\pi (\bx_T) 
\end{equation}
where $\log p_\pi (\bx_T) $ represents the final Gaussian noise distribution, ${s}_\theta (\bx(t),t)$ represents the time conditional score model at scale $t$, and $\nabla .$ is the divergence operator. This makes it computationally expensive to realize a MAP estimate using diffusion models.

\subsection{Architecture and implementation}
\subsubsection{DEEPEN}
For sampling, we used the same E2E trained network discussed in the previous section. Langevin algorithm in \eqref{langevin_update} was used to generate $100$ different samples. The algorithm was initialized with zero-mean Gaussian noise and ran for $100$ iterations. 
\subsubsection{Diffusion models}
A time-conditional score model was trained using the loss in \cite{yansong}, where the noise was added to the data according to the following perturbation kernel:
\begin {equation}\label{pert_score}
p_{0t}(\bx(t)|\bx(0))=\mathcal{N}(\bx(t);\bx(0),\dfrac{1}{2\log\Lambda}(\Lambda^{2t}-1)\bI)
\end{equation}
where $t$ is uniformly sampled over $[0,T]$, $\{\bx(t)\}_{t=0}^{T}$ represents the diffusion  process, $p_{0t}(\bx(t)|\bx(0))$ represents the transition probability from $\bx(0)$ to $\bx({t})$, and $\bx(0)$ are the training data samples. The architecture of the score model was chosen as a time-conditional DRUnet with $64$, $128$, $256$ and $512$ channels. We used the sampling algorithm proposed in \cite{dps} with $1000$ iterative steps to sample the posterior distribution. We note that DEEPEN requires fives times fewer number of parameters than the time-conditional score model.
\vspace{-4mm}
\subsection{Sampling illustration}
\begin{figure}[t!]
    \centering
    \begin{subfigure}[b]{0.4\textwidth}  
        \includegraphics[width=\linewidth]{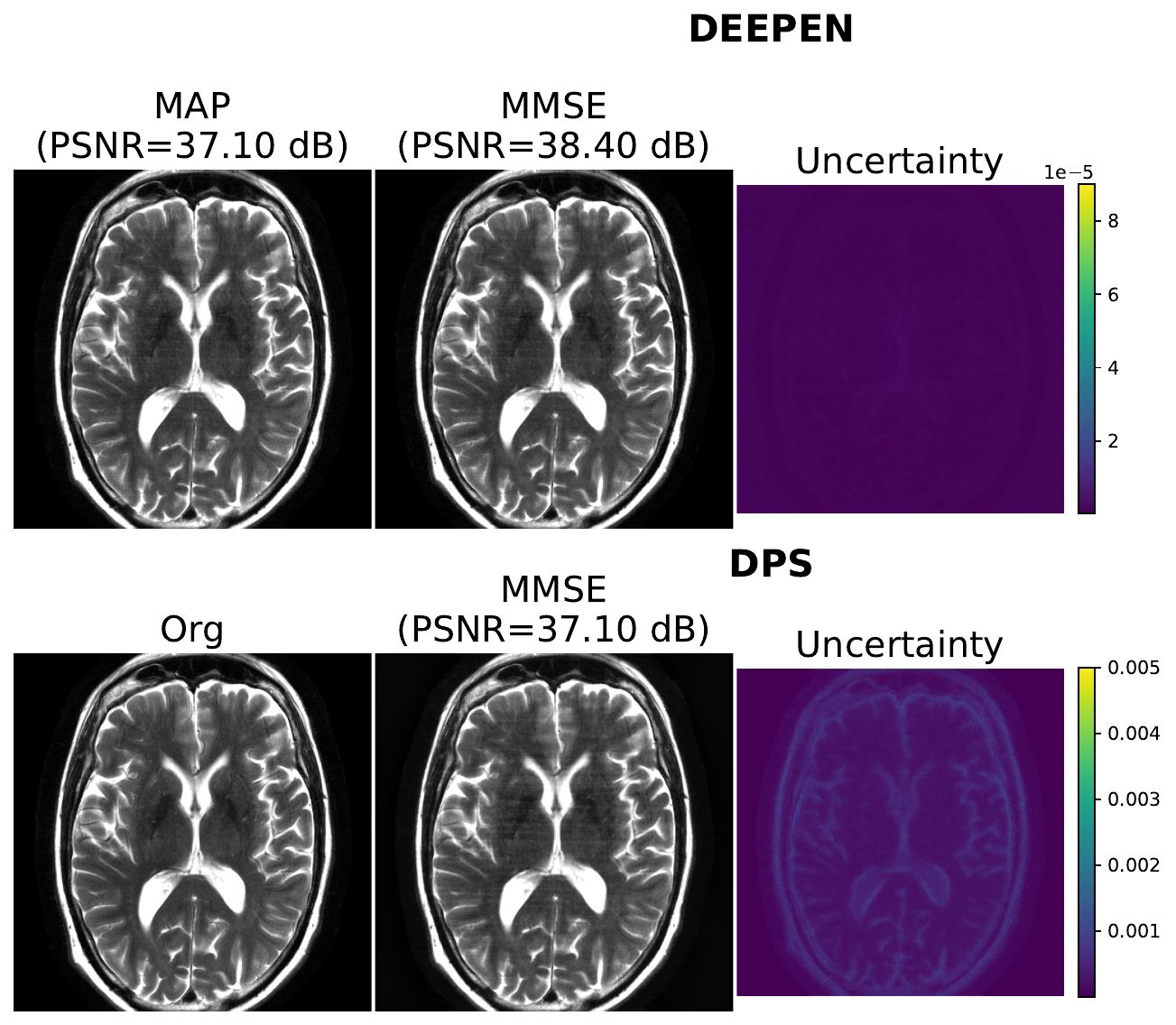}
    \caption{}
    \label{four_fold}
    \end{subfigure}
    \hfill
       \begin{subfigure}[b]{0.4\textwidth} 
        \includegraphics[width=\linewidth]{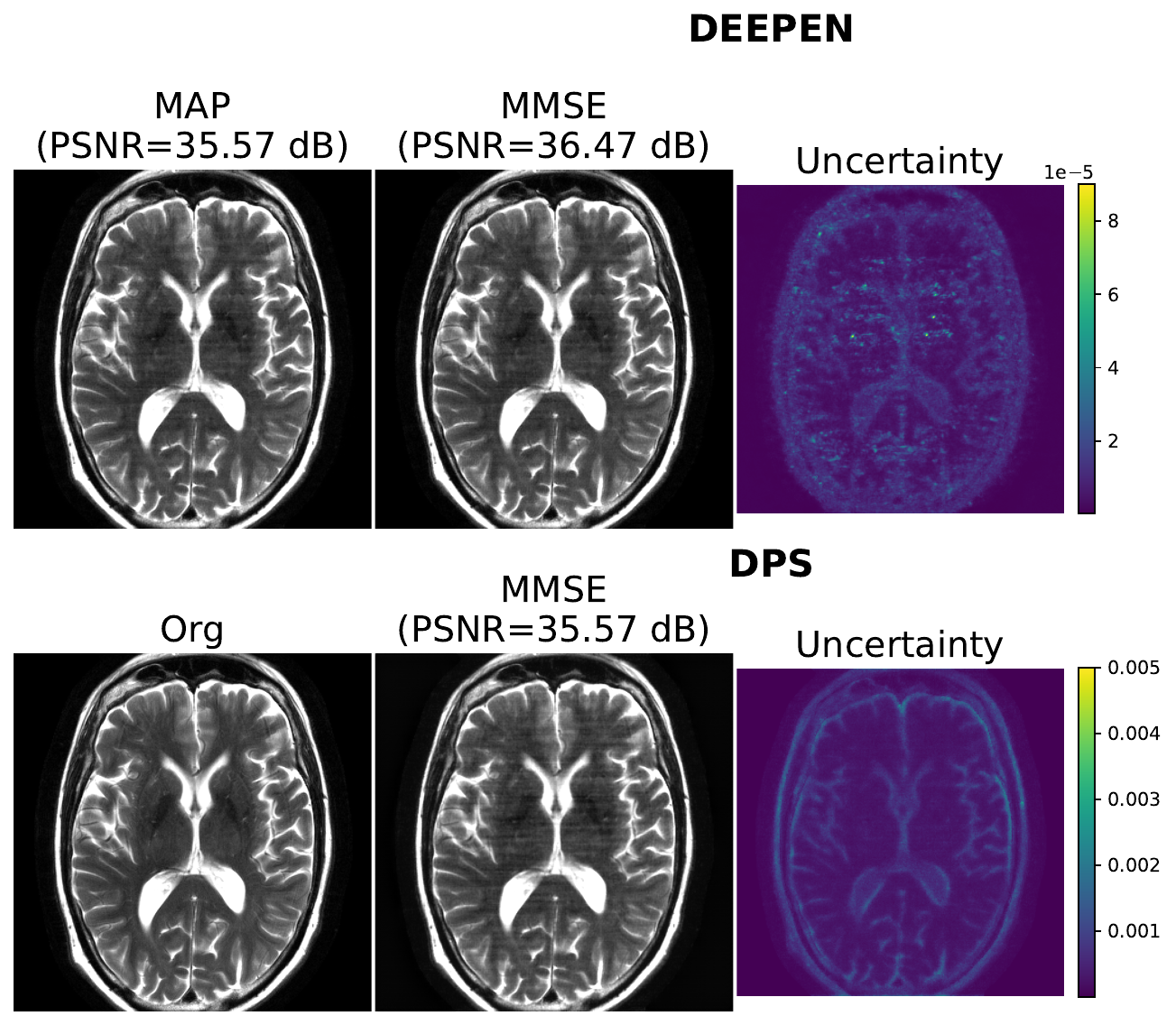}
    \caption{}
    \label{six_fold}
    \end{subfigure} 
    \caption{MAP, MMSE, and uncertainty estimate provided by DEEPEN algorithm for two different MRI acquisition settings. We compare DEEPEN's sampling performance with DPS when 2D undersampling mask is employed for (a) 4-fold and (b) 6-fold acceleration.\vspace{-7.5mm}} \label{sampling}
\end{figure}

Fig. \ref{sampling} illustrates the sampling performance of DEEPEN and DPS for different acquisition schemes. The upper row of each figure shows the MAP, the minimum MSE (MMSE), and the uncertainty estimates provided by the DEEPEN algorithm.  The MMSE and the uncertainty estimate are obtained by taking the mean and variance of the generated samples, respectively. The second row of each figure in Fig. \ref{sampling} shows the original image, MMSE, and the uncertainty estimates of the DPS algorithm. We do not show the MAP estimate as it is challenging to realize a MAP estimate using diffusion models, as discussed above.

\begin{table*}[]
\centering
\caption{Generalization performance of E2E-trained energy model: DEEPEN(boldfaced) and ELDER.}
\begin{tabular}{ll|ccc|}
\cline{3-5}
\multicolumn{2}{c|}{}                                                                                                               & \multicolumn{3}{c|}{Train setting}                                                                                                                                                            \\ \cline{3-5} 
\multicolumn{2}{c|}{}                                                                                                               & \multicolumn{1}{c|}{4x 2D mask}                                      & \multicolumn{1}{c|}{6x 2D mask}                                      & 2x 1D mask                                      \\ \hline
\multicolumn{1}{|l|}{\multirow{3}{*}{\begin{tabular}[c]{@{}l@{}}Test \\ \\ setting\end{tabular}}} & 4x 2D mask                      & \multicolumn{1}{c|}{\begin{tabular}[c]{@{}c@{}} $\mathbf{39.15+/- 1.43}$\\ $38.66 +/-1.23$\end{tabular}} & \multicolumn{1}{c|}{\begin{tabular}[c]{@{}c@{}}$\mathbf{37.60+/- 1.53}$\\ $38.88 +/-1.24$\end{tabular}} & \begin{tabular}[c]{@{}c@{}}$\mathbf{38.64+/-1.55}$\\ $37.86+/-1.20$\end{tabular} \\ \cline{2-5} 
\multicolumn{1}{|l|}{}                                                                            & 6x 2D mask                      & \multicolumn{1}{c|}{\begin{tabular}[c]{@{}c@{}}$\mathbf{37.50+/- 1.29}$\\ $36.82+/-1.18$\end{tabular}} & \multicolumn{1}{c|}{\begin{tabular}[c]{@{}c@{}}$\mathbf{37.18 +/-1.32}$\\ $37.39+/-1.17$\end{tabular}} & \begin{tabular}[c]{@{}c@{}}$\mathbf{37.20 +/-1.37}$\\$36.26+/-1.14$\end{tabular} \\ \cline{2-5} 
\multicolumn{1}{|l|}{}                                                                            & \multicolumn{1}{c|}{2x 1D mask} & \multicolumn{1}{c|}{\begin{tabular}[c]{@{}c@{}}$\mathbf{38.23+/- 1.99}$\\ $36.28+/1.43$\end{tabular}} & \multicolumn{1}{c|}{\begin{tabular}[c]{@{}c@{}}$\mathbf{37.66+/- 1.98}$\\ $36.54 +/-1.46$\end{tabular}} & \begin{tabular}[c]{@{}c@{}}$\mathbf{38.91+/- 2.21}$\\ $38.02+/-1.52$\end{tabular} \\ \hline
\end{tabular}
\label{generalization}
\end{table*} 

We note that DEEPEN requires 10x fewer sampling steps compared to DPS and uses a network with 5x fewer parameters, while offering comparable results.  
The narrower unimodal posterior distribution results in faster sampling and reduced uncertainty. This can be seen in Fig. \ref{sampling}.a and Fig. \ref{sampling}.b which shows the uncertainty estimates for a 4-fold 2D and 6-fold 2D undersampling mask, respectively. Deterministic ODE flows \cite{ddim,pnpflow} are designed to accelerate prior sampling by learning straighter paths between the source and target distributions. However, these models still require long chains for posterior sampling. A key challenge is that they are trained solely along the paths between the distributions, necessitating the addition of noise after each DC gradient descent step to project back to the originally trained regions \cite{pnpflow}. By contrast, DEEPEN gradients are efficient in removing both correlated and Gaussian perturbations, which we hypothesize to be another reason for the faster sampling.

We note that a benefit of the diffusion setting is that the same trained model can be re-used in multiple settings. In contrast, DEEPEN uses an E2E approach, which requires a customized model for each acquisition setting. However, the proposed E2E training strategy is more computationally and memory-efficient than traditional E2E models. In addition, the DEEPEN model is more generalizable than traditional E2E models to changes in the acquisition setting. Furthermore, it offers better performance and reduced uncertainty than diffusion models, eventhough it uses 10x fewer sampling steps and CNN models with 5x fewer parameters.
\vspace{-5mm}
\section{Conclusion}
We proposed DEEPEN, an E2E training framework to learn the posterior probability distribution for imaging inverse problems. DEEPEN can be used to derive the MAP estimate and sample from the posterior distribution. The proposed E2E model does not need to be unrolled, which results in a more memory-efficient training procedure. Unlike PnP and DEQ methods, this approach does not require a Lipschitz constraint for convergence, which translates to improved representation power and thus image quality. Our experiments show that the maximum likelihood training strategy offers a better-defined energy landscape, translating to improved image recovery algorithms. The experiments also show that the proposed approach can provide MAP estimates with improved image quality compared to E2E methods while being superior to PnP methods. We observe that the E2E approach can offer 10x faster sampling with models with reduced complexity compared to diffusion models, despite offering comparable reconstructions with reduced uncertainty. 
\vspace{-4mm}
\bibliographystyle{IEEEtran}
\bibliography{ref}

\end{document}